\documentclass[12pt,halfline,a4paper]{ouparticle}

\DeclareMathOperator{\erfc}{erfc}
\newcolumntype{P}[1]{>{\centering\arraybackslash}p{#1}}

\begin{document}

\title{An Abstract Model of Historical Processes}

\author{%
\thanks{Thanks to Oren Bernstein, Pavel Grinfeld, H{\aa}vard Hegre, Andrea Iff, Roger McCain, Donna Rizzo, Constantinos Rokas, and Stephen Wolfram for their feedback and suggestions.  All mistakes are my own.}
\name{Michael Poulshock}
\address{Drexel University, Thomas R. Kline School of Law}
}

\abstract{A theoretical model is presented which provides a way to simulate, at a very abstract level, power struggles in the social world.  In the model, agents can benefit or harm each other, to varying degrees and with differing levels of influence.  The agents interact over time, using the power they have to try to get more of it, while being constrained in their strategic choices by social inertia.  The outcomes of the model are probabilistic.  More research is needed to determine whether the model has any empirical validity.}


\maketitle

\section{Introduction and Motivation}

When we look back over the history of humanity, regardless of the time period, place, or culture that we examine, we find agents engaged in power struggles.  These agents might be states jockeying for primacy in the international arena.  They might be groups or factions grappling for control of some market or political apparatus.  They might be individuals competing for supremacy within some institution or spatial region.  At all of these various levels, across all of history, the existence of power struggles like these has been invariant, presumably a function of the fact that throughout all of human history, \textit{there have been humans present}.

What do we mean by \textit{power struggles}?  Simply put, they are conflicts that arise due to the tendency of agents to use whatever power they have to get \textit{more} of it.  In other words, they are struggles over power itself.  When struggling for power, agents must decide who to cooperate with, and whom to harm, allocating whatever power is at their disposal and taking into consideration the anticipated actions and reactions of their competitors.  They tend to seek a combination of both absolute and relative power.  The political theorist Hans Morgenthau (1954) had this sort of process in mind when he described power, in the international context, as ``that untamed and barbaric force which finds its laws in nothing but its own strength and its sole justification in its own aggrandizement."  

This paper presents a model of power struggles.  It ignores many of the things that affect real world conflict dynamics, like culture, technology, the environment, ideology, geography, migration, institutions, disease, and resource scarcity.   Instead, it focuses on the fluid dynamics of a single thing: power.  It seeks to answer the question: Given an initial configuration of relationships among agents, each of whom has a starting amount of power, how will the system tend to evolve in time?  Or: If all we knew about a social situation were the relative powers of the actors and the network topology of the alliance structure, to what degree could we anticipate future states, given some minimal set of assumptions? The model illustrates how the struggle for power in and of itself might give rise to conflict and alliance dynamics that resemble those that we see in the real world.  Future work will refine the model and attempt to apply it to historical situations in order to gauge its empirical utility; at the moment, it is merely hypothetical.

\section{Foundational Ideas}

We'll begin by stating the foundational ideas underlying the model.  First, some definitions:  An \textit{agent} is an individual actor or a collection of individuals acting in concert, such as a civilization, nation, firm, etc.  \textit{Power} is a quantity reflecting an agent's ability to affect the power of other agents.  \textit{Benevolence} is an agent's expenditure of power that increases the power of another agent.  \textit{Malevolence} is an agent's expenditure of power that decreases the power of another agent.  And a \textit{tactic} is the way an agent allocates its power, both benevolently and malevolently, towards other agents.

We postulate the following:

\begin{enumerate}
  \item Reciprocal benevolence increases the power of each agent.
  \item Power used malevolently has more impact than power used benevolently.
  \item Agents seek power in both an absolute and relative sense: they want to increase the amount of power that they have, but they also want more power than others.
  \item Agents interact indefinitely, but at any point in time they must make tactical decisions in the face of (inherent, exogenously generated) uncertainty about the future. 
  \item The larger the change in an agent's tactics, the less likely it is, due to \textit{social inertia}.
  \item Agents are rational in the sense that they act in their own self-interest and assume that others do likewise.
\end{enumerate}
In this paper, we make the additional assumption that agents have perfect information: they each know everything that every other agent knows.  

We will use these notions, discussed in more detail below, to generate simulations. The idea behind a simulation is that you start with some initial configuration or state, along with rules that alter the state, and then you repeatedly apply those rules and see what happens. Here, there are two distinct layers of simulation that occur.  At a high level, we will generate a series of \textit{frames}, or states, that can be played in sequence like a film.  Our ultimate goal is to watch those films.  But in order to generate each individual frame, a separate, lower-level simulation must occur to explore possible futures at that particular point in time.  We will describe the lower-level simulation and then explain how results from it are composed to form the higher-level one.  But we first need to describe the core abstraction of the model, which defines how agents interact with each other and what they want.

\section{The Core Model}

The model is presented here in mathematical form.  A source code implementation in the \textit{Mathematica} programming language is available at \url{www.github.com/mpoulshock/AMHP}. 

\subsection{Basic mechanism}

We use a network, or graph, structure to model agents and their interrelationships.  Each node, or agent, has a characteristic called \textit{power}, which quantifies its ability to influence events.  Specifically, it is an agent's ability to alter other agents' power, a deliberately recursive definition. Power is also referred to as \textit{size} and represented as a number $s$, where it is always the case that
\begin{equation}\label{eq:1}
s \ge 0
\end{equation}
The larger the number, the greater the agent's power.  Zero power means that the agent has no influence, even over itself, and is effectively dead.

An agent can direct its power positively or negatively towards other agents.  Using it positively, or benevolently, entails giving it to another agent.  However, when one agent gives a positive amount of power to another, the amount received by the other agent is increased by a factor of $\beta$, where
\begin{equation}\label{eq:2}
\beta > 1
\end{equation}
This \textit{benevolence multiplier} reflects the idea that, regardless of what has been given, it has more value to the recipient than it did to the giver.  This derives from postulate (1): if reciprocal benevolence increases the power of each party, then unilateral benevolence increases the power of the recipient.  The increase factor can be thought of as possibly corresponding in the real world to the benefits of exchange and the division of labor.  For example, if S sells B a pound of apples for \$1, we can assume that to B the apples were more valuable than \$1, and that to S the \$1 was more valuable than apples.  The benevolence multiplier allows agents in the model to achieve mutual benefit (growth in power) by cooperating with each other.  Power used towards benevolence is expressed as a positive number.

The transfer of negative power, or malevolence, is similarly subject to a \textit{malevolence multiplier}, $\mu$, where
\begin{equation}\label{eq:3}
\mu > \beta
\end{equation}
That is, when an agent uses power negatively, for every unit of power it expends, it causes a reduction in the recipient's power by $\mu$.  This captures the idea that it is easier to destroy value than to create it, as expressed in postulate (2) above.  Exactly how much easier is left undefined, as one of the model's parameters.  To the extent that this model has real world correspondence, it may be that an empirical value for $\mu$ bears some relationship to the concept of entropy, since causing harm or destruction is ultimately about increasing disorder.  Power used towards malevolence is expressed as a negative number.

An agent can never use more power than it has.  That is, the sum of the absolute value of an agent's outgoing allocations to other agents must equal the agent's total power.  A \textit{tactic vector} $\tau$ represents an agent's allocation of its power, where for each element
\begin{equation}\label{eq:4}
\tau_{j} \in [-1,1]
\end{equation}
and
\begin{equation}\label{eq:5}
\sum_{j=1}^{n} \left|\tau_{j}\right| = 1
\end{equation}
where $n$ is the number of agents.  The tactic vector expresses an agent's relationships as positive and negative percentages whose absolute values must sum to one.  It can be thought of as a kind of a foreign policy describing the agent's conduct towards other agents.  An example of a legal tactic vector where there are three agents is
\[ \left( 0.7,-0.1,0.2 \right) \]
The $j$th index of a tactic vector represents the percentage of its power that an agent allocates towards itself. This is not subject to any multiplier. 

All of the agents' tactic vectors together form a \textit{tactic matrix}, $\mathbf{T}$.  Here's an example:

\[ \left( \begin{array}{ccc}
0.7 & 0.1 & 0 \\
-0.1 & 0.8 & 0 \\
0.2 & 0.1 & 1 \end{array} \right)\] 
Each column in this matrix represents a tactic vector and satisfies equation (\ref{eq:5}).  The matrix diagonal represents the amount of power that the agents are allocating to themselves.

The agents' sizes are stored in a \textit{size vector}, $\mathbf{s}$, for example

\[ \left( 0.3,1,0.6 \right) \]
To ensure consistent results from one simulation run to the next, we'll adopt the convention of normalizing the initial tactic vector so that the largest agent has a size of 1.

The tactic matrix and the size vector together comprise the \textit{state}. The state can be visualized as a graph in which the node sizes show the power of the agents and the directed edges, varying in thickness, show the agents' tactical allocations.  Figure 1 shows an example using the size vector and tactic matrix above.  Green represents positive (benevolent) transfers and red negative (malevolent) ones.

\begin{figure}[!htbp]
\centering
\includegraphics[width=5cm]{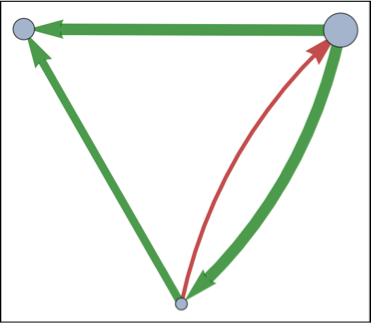}
\caption{State graph}
\end{figure}

The state is updated in time steps, with all agents moving simultaneously.  At each step, power is transferred among the agents according to their tactic vectors and as it flows around the network, the power of each agent increases or decreases.  An agent's updated power is the sum of the power coming to it from other agents, subject to the appropriate multiplier, plus the power that it kept to itself.  We can represent this updating procedure mathematically as

\begin{equation}\label{eq:6}
\mathbf{s}(t) = (\mathbf{T} \circ \mathbf{M})^t \, \mathbf{s}(0)
\end{equation}
where $\circ$ represents the Hadamard product (element-wise matrix multiplication) and $\mathbf{M}$ is a matrix representing the benevolence and malevolence multipliers defined as 

\begin{equation}\label{eq:7}
(m_{ij}) = \begin{cases} 
      1 & i = j \\
      \beta & \tau_{ij} \ge 0 \\
      \mu & \tau_{ij} < 0 \\
   \end{cases} 
\end{equation}
Another way to express equation (6), using the prior state rather than the initial state, is
\[
\mathbf{s}(t+1) = (\mathbf{T} \circ \mathbf{M}) \, \mathbf{s}(t)
\]
It's important to add that in equation (\ref{eq:6}), for every element of $\mathbf{s}$, $s_{i} \ge 0$, due to the constraint imposed by equation (\ref{eq:1}).  If, as a result of updating, an agent's size becomes less than or equal to zero, they are considered dead and assigned a size of zero.  The existence of negative sizes, besides being meaningless, would cause subsequent update operations to behave counterintuitively.

Using the example data above, and assuming that $\beta=1.2$ and $\mu=3$, the new size vector produced by equation (\ref{eq:6}) would be

\[ 
\left( 
\left( \begin{array}{ccc}
0.7 & 0.1 & 0 \\
-0.1 & 0.8 & 0 \\
0.2 & 0.1 & 1 \end{array} 
\right) \circ \left( \begin{array}{ccc}
1 & 1.2 & 1.2 \\
3 & 1 & 1.2 \\
1.2 & 1.2 & 1 \end{array} 
\right)
\right)^1 \cdot 
\begin{pmatrix}0.3\\1\\0.6\end{pmatrix} = 
\begin{pmatrix}0.33\\0.71\\0.79\end{pmatrix}
\]

Though abstract, the formalism of this model provides a flexible way of expressing a variety of interrelationships.  Actions can be constructive or destructive, and in varying degrees.  Relationships need not be symmetrical, as they frequently are not in real life.  And some agents have more influence than others.  A variety of situations can be modeled this way.

We have described the model's core data structures and how the state is updated from one time step to the next.  Now we turn to the question of what agents seek.

\subsection{Utility}

At each time step, each agent can adjust its tactic vector.  These choices can lead to a wide variety of possible states, and agents need a way to evaluate whether a given state is better or worse for them than any other state.  So what do agents want?  We define three notions of utility that will reflect agents' preferences.  Later, we will use these various utility functions to evaluate possible moves.

\subsubsection{Positional Utility}

The first utility function, \textit{positional utility}, defines the way that agents seek to maximize power.  They do not merely want to maximize their own power.  They instead maximize a utility function approximated by

\[
u_{i} \approx \frac{s_{i}^{2.5}}{\displaystyle\sum_{j=1}^{n} s_{j}^2}
\]
where $u_{i}$ is the utility to the $i$th agent.

Basically, two things matter to agents: they want to be powerful in an absolute sense, and they want to dominate other agents by being powerful in a relative sense.  These two desires are in tension: to dominate, they may have to shrink by causing others to shrink even more; to grow in size, they may have to cease dominating.  The utility function encodes these conflicting objectives.  This is easier to see when the right side of the utility function is split into two components:

\[
u_{i} \approx \frac{s_{i}^2}{\displaystyle\sum_{j=1}^{n} s_{j}^2} \sqrt{s_{i}}
\]

The first component on the right side of this equation reflects dominance: specifically, the ratio of an agent's \textit{market share} (their proportion of all power in the system) squared to the total of each agent's market share squared.  In economic terms, this expresses each agent's contribution to the Herfindahl-Hirschman Index, which is a measure of competition within a given market.  This component embodies the idea that the smaller and more divided one's competition, the better off one is.  (By way of comparison, John Mearsheimer (2001) asserts that states in the international system seek to maximize their \textit{market share} of power.)

The second component of the equation above provides an incentive for absolute growth.  The square root of size is taken in order to reflect the diminishing marginal utility of acquiring power, since one additional unit of power is more valuable to a small agent than to a large one.  (This is a common facet of economic utility functions.)  This could be generalized to a cube or $n$th root, such that

\begin{equation}\label{eq:8}
u_{i} = \frac{s_{i}^\alpha}{\displaystyle\sum_{j=1}^{n} s_{j}^2}
\end{equation}
where the utility exponent $\alpha$ is in the range
\begin{equation}\label{eq:9}
\alpha \in [2,3]
\end{equation}
As $\alpha$ decreases, relative power is incentivized and agents become more apt to use violence to cut other agents down to size, so they can hoard market share.  As $\alpha$ increases, they're more prone to pursue absolute growth, which requires cooperation (mutual benevolence).  We could allow $\alpha$ to vary for each individual, which might better reflect the heterogeneity of the social world.  However, for the sake of simplicity, we'll assume that this parameter is the same for all agents.

This utility function has a few other desirable properties worth mentioning.  First, agents that are the same size have the same payoff, dead agents have a payoff of zero, and the largest agents will have the largest payoff.  Adding agents with a size of zero to the population doesn't affect the payoffs to the existing agents.  Further, when there are two agents whose sizes differ by a constant amount, their payoffs will tend to converge as their sizes increase by the same amount (e.g. the payoffs to agents with $\mathbf{s} = (100,101)$ will be closer together than those with  $\mathbf{s} = (1,2)$).  Finally, the utility function is smooth and well-behaved, except when all agents have a size of zero, in which case there's no one left to care.

\subsubsection{Expected Utility}

It cannot be the case that agents are free to choose \textit{whatever} tactics they like.  In the real world, the past, or more specifically the present, binds their options.  For example, a country could not one day suddenly alter its entire foreign policy, even if taken over by a madman.  Processes like trade agreements, peace talks, mergers and acquisitions, and divorces all take time, because social relationships have a kind of stickiness that resists rapid change.  This phenomenon is called \textit{social inertia} (Wikipedia 2016), and it makes it less likely that agents will be able to effect dramatic tactical changes (postulate (5)).  Faraway tactics are simply unlikely to be reached; ones more similar to the current tactic are more plausible.  

So the rewards that an agent would ordinarily get from its positional utility have to be reduced if those positions are the result of unlikely tactical changes.  Following a convention in economic modelling, we'll let the expected value of an event equal its value times the probability of it happening.  And we'll assume that all agents assume that all other agents assume this; so there is communal acknowledgement of the external constraint posed by postulate (5) and its effect on everyone's rewards.

Accordingly, an agent's \textit{expected utility} $p$ from a position will be the position's utility to the agent times the probability of getting to the current tactic matrix $\mathbf{T}_{t}$ from the previous one $\mathbf{T}_{t-1}$:

\begin{equation}\label{eq:10}
p_{i} = u_{i} \, q(||\mathbf{T}_{t} - \mathbf{T}_{t-1}||_{F})
\end{equation}
Here $|| \cdot ||_{F}$ is the Frobenius norm, a measure of the Euclidean distance between the two tactic matrices, and
\begin{equation}\label{eq:12}
q(x) = \erfc(\frac{x}{\sigma \sqrt{2}})
\end{equation}
The $q$ function approximates the likelihood of the tactical distance being traversed, and therefore of the new position being reached.  This function gives the tail probability of the half-normal distribution, just as the statistical Q-function gives the tail probability of the normal distribution.  In a half-normal distribution, only nonnegative values of x are possible, and it's used here because tactical distance is never negative.  (The \textit{erfc} function is the complementary error function.)  The variable $\sigma$ is used as a model parameter that controls the intensity of social inertia: when $\sigma$ is close to zero, inertia is high; as $\sigma$ increases, inertia decreases.

What this means is that the greater the tactical distance to a state from the prior one, the less likely that state is to be reached.  And that likelihood is based on the shape of a bell curve, so that shorter distances are much more likely than longer ones.  The shape of the bell curve, and therefore the intensity of social inertia in the model, can be adjusted by altering $\sigma$.

Summarizing, expected utility is the power that an agent can expect to have at a particular future state, taking into account the likelihood of that state actually being reached in a world afflicted by social inertia.

\subsubsection{Intertemporal Utility}

The third type of utility is \textit{intertemporal utility}.  It takes into account the fact that the interaction among the agents goes on indefinitely and that at any given moment a cloud of uncertainty hangs over the future, per postulate (4).  The fact that there is a future at all is significant, as it opens up the possibility of reciprocity because agents can expect future rewards. The fact that the future is uncertain means that rewards in the near term are more valuable than those farther in the future. 

We define the intertemporal utility $P$ of a strategy as the total rewards that an agent will receive over future time steps, subject to discounting such that rewards in the near future are weighted more heavily than those in the distant future:
\begin{equation}\label{eq:13}
P_{i} = (1-\delta) \sum_{t=1}^{\infty} \delta^t p_{i}(t)
\end{equation}
where $\delta$ is the \textit{discount factor} and
\begin{equation}\label{eq:14}
\delta \in (0,1)
\end{equation}
When $\delta$ increases, future rewards are given more weight and agents can be thought of as having more patience, enabling greater cooperation.  Conversely, when $\delta$ decreases, agents are more short-sighted and eager for immediate rewards.  The term $(1-\delta)$ normalizes the result so it can be compared with the payoffs from individual time steps. (Ratliff 1996)

Intertemporal utility, then, tells agents what payoffs they will receive from a sequence of moves, not just from the next immediate move.  It allows agents to take the future into their strategic calculations and to consider various long-term scenarios and decide which they prefer.

\section{Generating Simulations}

Thus far, we have described the mechanics of how agents interact and the nature of the rewards that they seek.  In this section, we describe how those rules give rise to a dynamic simulation that unfolds over discrete time steps.  We will first explain the lower-level simulation that generates individual frames, before moving on to define \textit{reels}, which are the high-level simulations that we are interested in observing.  An unusual feature of this simulation concept is that for any initial state, we can examine numerous possible futures and their associated probabilities.

\subsection{Generating Frames}

From any given state, we need a way of determining the next state.  In the real world, strategic agents consider their own possible moves and those of their competitors, and engage in an \textit{if I do that, she'll do that...}kind of exercise.  Similar to the way a chess player imagines various game permutations before making an actual move, the agents in this model engage in speculation about what to do before actually moving on to the next frame.  With our bird's eye view of the system, we will perform a Monte Carlo simulation, generating numerous random sequences of tactic matrices, or \textit{lines of play}, and then eliminating the lines which are \textit{irrational}.  We use the remaining, rational lines of play to determine what the agents are likely to do on their next move, and we end up with a set of possible next frames and associated probabilities for each.

\subsubsection{Identifying Rational Lines of Play}

We have asserted that agents are rational (postulate (6)) and we will appeal to a notion of rationality established in the field of game theory.  The idea is that agents will take future-oriented intertemporal rewards if they are better than what each agent is guaranteed to get if there were no future and all that mattered were the expected utility of the next immediate move.  Game theorists typically illustrate this idea using the Prisoner's Dilemma, a toy game in which the players' best responses result in an apparently suboptimal outcome because there is no incentive for cooperation.  In contrast, when the Prisoner's Dilemma is played repeatedly, cooperation becomes possible because players are able to anticipate future rewards and punishments (see Ratliff 1996).  We will use this concept to identify rational lines of play among our agents.

We proceed in two stages.  First, we identify the best outcome that each agent is guaranteed to receive on the next move.  This is known as the \textit{minimax vector}.  To compute it, we select random possible tactics for each agent and then identify each agent's best responses to the other agents' tactics.  Here, the best response is the one that results in the highest expected utility $p$.  If there are any tactic matrices that comprise all agents' best responses, these are the \textit{stage Nash equilibria}.  There can be one, more than one, or sometimes zero such equilibria.  If there are more than one, we look for each agent's lowest expected utility among the equilibria.  The result of this first step, the minimax vector, is a size vector representing the minimum power each agent is guaranteed to have on the next move.

\begin{figure}[!htbp]
\centering
\includegraphics[width=13cm]{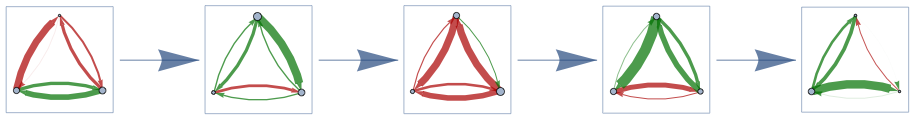}
\caption{A line of play is a sequence of states generated from a list of random tactic matrices. At each step, agents' sizes are updated by applying equation (6).}
\end{figure}

We then look for lines of play that make everyone better off than the stage minimax vector.  For a particular line of play, we generate a sequence of random tactic matrices, play out each of them, and compute the intertemporal utility $P$ of the sequence.  Any line of play in which an element of the intertemporal utility vector is less than the corresponding element of the minimax vector can be eliminated as irrational.  No rational agent would play their respective moves in that line because they'd do worse than they would by playing a move that leads to a stage Nash equilibrium.  In game theory, this idea is known as the \textit{folk theorem}.  We want to keep only those lines of play where, for each agent
\begin{equation}
P_{i} > minimax_{i}
\end{equation}
After eliminating the irrational lines of play, we are left with those that are plausible, in the sense that rational agents would select them over the short-term stage Nash moves.  In other words, we have found sequences of moves in which the first move of each sequence is a plausible candidate for the next frame.

\subsubsection{Determining the Likelihood of Various Next States}

We have identified rational lines of play and in each, the first move is a contender for being the next frame.  We now want to know the likelihood of each of these contenders.

We can determine the likelihood of a line of play based on the total tactical distance that it covers.  Due to social inertia, the longer the distance, the less likely the line of play.  And the probability of a line of play is the probability that the first move along it will, in fact, be played.  The idea is that agents contemplate their various options and then must choose their next step into the haze of the future.  We define the weight $w$ of a next frame as being inversely proportional to the total (intertemporally discounted) distance of its line of play:
\begin{equation}
w(\mathbf{T}_{1}) = q \big( (1-\delta) \sum_{t=1}^{\infty} \delta^{t} \, ||\mathbf{T}_{t} - \mathbf{T}_{t-1}||_{F} \big)
\end{equation}
This equation follows a similar pattern as (12).  The weightier a frame, the more plausible it is.

Depending on the number of lines of play that we have randomly generated, there could be many rational first moves (tactic matrices).  We can bring some coherence to this mass of information by clustering similar tactic matrices together.  There are quite a few ways to do this, but perhaps the simplest is to round the elements of each tactic matrix to a desired multiple.  

Clustering allows us to accumulate the overall probability of each possible next frame.  We define the probability $\mathbb{P}$ of transitioning from $\mathbf{T}_{0}$ to a possible next frame $\mathbf{T}_{1,a}$ as the total of all weights associated with $\mathbf{T}_{1,a}$, divided by the total weights for all next frames that were generated:
\begin{equation}
\mathbb{P}(\mathbf{T}_{0}|\mathbf{T}_{1,a}) = \frac{\sum w(\mathbf{T}_{1,a})}{\sum w(\mathbf{T}_{1, \cdot})}
\end{equation}
The more frequently a frame appears in the sample of Monte Carlo-generated (rational) lines of play, the more likely it is.  And the greater its weight, the more likely it is.  As more lines of play are randomly generated, the probability estimates for the various possible frames converge to stable values.

To recap, we have explained how to produce a frame - actually, numerous possible frames - from a prior one.  First, we generate random lines of play (sequences of tactic matrices) into the imagined future.  Of these, we select those that are rational, in the sense that the long-term rewards to each agent surpass any guaranteed short-term gain.  We then estimate the likelihood of each line of play based on its tactical variation, and finally we cluster the first states on those paths to estimate the likelihood that each of those states will be the next frame.

This simulation concept is probabilistic: state A will not invariably give rise to state B; instead, the model yields a probability of transitioning from A to B.  So from any starting frame, we end up with a spectrum of possible subsequent frames.

\subsection{Generating Reels}

A reel is a series of frames that can be played like a film.  We can generate a reel by repeating the process described in section 4.1 for every new frame.  Starting from an initial state, we build a tree of possible next frames, applying the above procedure to every leaf state, going as deep into the future as we like.  This process generates a chain of states and, between any two states, the likelihood of transitioning from one to the next.  We end up with a probability tree where each path from the origin is a reel (see Figure 3).  The overall structure is not a single reel, but a family of reels all spinning out from the initial state.  This tree structure allows us to see a variety of possible futures and to gauge the ultimate likelihood of each.

\begin{figure}[!htbp]
\centering
\includegraphics[width=14cm]{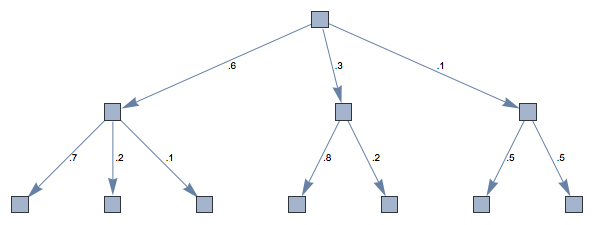}
\caption{Each square is a state and edges show the probabilities for transitioning from one state to the next. Each path from the initial state (top) to a leaf state is a reel.  The transition probabilities are computed from the lower-level simulation using equation (16).}
\end{figure}

We can calculate the likelihood that a particular reel \textbf{R} will occur, given an initial state, as 
\begin{equation}
\mathbb{P}(\mathbf{R})  = \prod_{t=0}^{t_{max}} \mathbb{P}(\mathbf{T}_{t}|\mathbf{T}_{t+1})
\end{equation}
Here we multiply the probabilities of each frame transition along the reel, following the multiplication rule for independent events.  For example, in Figure 3, the likelihood of the rightmost path occurring would be $.1 \times .5 = .05$.

Probabilistic outcomes are important because the social world is non-deterministic.  Any model that asserted that a particular outcome was the only possible one would be implausible.  Here, the probabilistic nature of the outcomes is not the result of the random nature of the Monte Carlo simulation; instead it arises from postulate (5), which provides a way to gauge the likelihood of events by the degree to which they breach social inertia.

\section{Discussion and Future Work}

By exploring the implications of some fairly straightforward foundational ideas, we have defined a probabilistic simulation model, represented more or less by the numbered equations above, in which agents capable of inflicting harm on others seek to maximize their own power.  This model provides a framework for reasoning about power struggles and possibly for projecting their outcomes.

Does the real world work this way?  Certainly, the basic assumptions of this model align with intuitive notions of power and its pursuit. Historical agents continually rearrange their relationships in order to prosper, dominate, and avoid being dominated, and these rearrangements simultaneously cause changes to the wider network and are reactions to it.  The universality of power struggles at many levels of the social world suggests that they may be generated by a single underlying process.  So it doesn't seem insane to wonder whether the model has any empirical merit.

To test it, we would need to initialize simulation runs with historical scenarios and see whether the outcomes resemble reality.  We could also test the model on patterns commonly seen in power struggles and assess whether the outcomes meet expectations for how those situations typically unfold.  This would require tinkering with the model parameters ($\alpha$, $\beta$, $\mu$, $\delta$, and $\sigma$) to see what ranges lead to the most plausible behavior. We should not expect perfect fidelity, as the model's simplistic assumptions - perfect information, perfect rationality, perfect computational ability, and perfectly homogeneous agents - obviously do not hold in the real world.  Several aspects of this experimental effort merit comment.

First, regarding the subjects of our investigations, one would expect that agents that are aggregates of many individuals - for example, nations - would be more likely to conform to the idealization of the model.  The same holds true of power struggles that develop over longer time scales.  We also have to be wary about applying the model to historical eras in whose interpretation we have a vested interest.  The farther we look from our own historical myths and ideologies, the more objectively we can appreciate the role that raw power plays in history.

Second, quite a lot of computation is required as the number of agents increases and as the horizon of the future lengthens.  For nontrivial historical situations, it may prove difficult to scale this model up.

Third, we need to determine whether and how \textit{power} in the model is reflected in measurable real world quantities.  The definition has to permit a fungibility between benevolence and malevolence: capital needs to be convertible into violence and vice versa. In the international context, many measures of national power are available (H{\"o}hn 2011).

Finally, we might consider extending the model to include other variables, like resource scarcity, the location and movement of agents in space, the role of institutions in constraining and channeling power, and information asymmetries, the last of these being of particular interest as a relaxation of the assumption we made above pertaining to perfect information.  

\section{Data structures and variables}

For convenience, we list the model's core data structures and variables.  

\vspace{5mm}
\begin{tabular}{ |P{1.5cm}|p{4cm}| }
\hline
\textbf{Symbol} & \textbf{Meaning} \\
\hline
$n$ & Number of agents \\
$\mathbf{s}$ & Size (power) vector \\
$\mathbf{\tau},\mathbf{T}$ & Tactic vector, matrix \\
$\mathbf{M}$ & Multiplier matrix \\
$t$ & Time step \\

\hline
\end{tabular}
\vspace{5mm}

Below are parameters that can be tuned when searching for correspondence with real world behavior.  These five parameters are related to the first five postulates.

\vspace{5mm}
\begin{tabular}{ |P{1.5cm}|p{5cm}|P{1.5cm}|P{2cm}| }
\hline
\textbf{Symbol} & \textbf{Meaning} & \textbf{Range} & \textbf{Postulate} \\
\hline
$\beta$ & Benevolence multiplier & (1, $\mu$) & 1 \\
$\mu$ & Malevolence multiplier & ($\beta$, $\infty$) & 2 \\
$\alpha$ & Positional utility exponent & [2, 3] & 3\\
$\delta$ & Discount factor & (0, 1) & 4 \\
$\sigma$ & Coefficient of social inertia & (0, $\infty$) & 5 \\

\hline
\end{tabular}
\vspace{5mm}

\section{References}

\begin{enumerate}

\item H{\"o}hn, Karl Hermann, \textit{Geopolitics and the Measurement of National Power} (dissertation), available at \url{http://ediss.sub.uni-hamburg.de/volltexte/2014/6550/pdf/Dissertation.pdf} (accessed Dec. 12, 2016).

\item Mearsheimer, John J. (2001). \textit{The Tragedy of Great Power Politics}. New York, NY: W.W. Norton \& Company, Inc.

\item Morgenthau, Hans (1954).  \textit{Politics Among Nations: The Struggle for Power and Peace}.  New York, NY: Alfred A. Knopf.

\item Ratliff, Jim (1996). \textit{Infinitely Repeated Games with Discounting}, available at  \url{http://www.virtualperfection.com/gametheory/5.2.InfinitelyRepeatedGames.1.0.pdf} (accessed Dec. 12, 2016).

\item Wikipedia (2016). \textit{Social Inertia} entry, available at \url{https://en.wikipedia.org/wiki/Social_inertia} (accessed Dec. 12, 2016).

\end{enumerate}

\end{document}